\documentclass{article}

\begin{document}

\def \be {\begin{equation}}
\def \ba {\begin{eqnarray}}
\def \ee {\end{equation}}
\def \ea {\end{eqnarray}}
\def \lsm {L$\sigma$M}
\def\bea{\begin{eqnarray}}
\def\eea{\end{eqnarray}}
\def\lsim{\;\raise0.3ex\hbox{$<$\kern-0.75em\raise-1.1ex\hbox{$\sim$}}\;}
\def\gsim{\raise0.3ex\hbox{$>$\kern-0.75em\raise-1.1ex\hbox{$\sim$}}}

\author{Nils A. T\"ornqvist \\
Department of Physical Sciences, \\ University of Helsinki, POB
64, FIN--00014}
\title{Mixing the Strong and E-W Higgs Sectors.\footnote{To appear in Physics Letters B}}

\maketitle

\begin{abstract}
After noting the well known similarity of the minimal electro-weak
Higgs sector with the  linear sigma model for the pion and the
sigma, it is found that a small mixing term between the two models
generates a pion mass. Although the  custodial $SU(2)_L\times
SU(2)_R$, and the gauged $SU(2)_L\times U(1)$ symmetry for the
whole model remains intact, the mixing breaks the relative chiral
symmetry between the two sectors. The mixing should be calculable
from light quark masses as a quantum correction. This simple
mechanism of "relative symmetry breaking" is believed to have
applications for other forms of symmetry
breaking. \\
\noindent Pacs numbers: 11.15.Ex, 12.39.Fe, 14.80.Bn

\end{abstract}
\medskip

Spontaneous symmetry breaking in the vacuum is certainly a very
beautiful and important concept in many areas of physics. The
prototype for this mechanism in particle physics is given by the
linear sigma model (\lsm)\cite{LSM}, (which can be looked upon as
an effective theory at sufficiently low energy of a more
fundamental theory as QCD). The minimal electro-weak scalar sector
was essentially copied from the \lsm, except for a much higher
value for the vacuum expectation value, $v=246$ GeV, instead of
the  $92.6$ MeV in the \lsm, which we here denote by $\hat v$. The
latter value, $\hat v$, is also the pion decay constant ($f_\pi$),
and is  proportional to the $q\overline{q}$ condensate of QCD.
Thus the vacuum values are orders of magnitudes different, $v/\hat
v$ is about 2656.

The analogy is more evident if we represent the conventional
complex Higgs-doublet composed of $\phi^{+}$ and $\phi^{0}$ by a
seemingly redundant matrix form (We follow the notations of
Willenbrock\cite{Willen})

\be\left(
\begin{array}{c}
 \phi^{+} \\
\phi^{0} \\
        \end{array}\right ) \to
\frac 1 {\sqrt 2} \left(
\begin{array}{cc}
\phi^{0*} & \phi^{+}  \\
-\phi^{-}& \phi^{0} \\
        \end{array}\right )  =\Phi .   \ee
The  Higgs Lagrangian then takes the form

\be
 {\cal L_{\it{Higgs}}}(\Phi ) =  {\rm Tr}
[(D_\mu\Phi )^\dagger D_\mu\Phi] + \mu^2{\rm Tr} [\Phi^\dagger
\Phi] -\lambda {\rm Tr}[\Phi^\dagger\Phi ]^2, \label{higgs}\ee
from which  (at the tree level) the vacuum value is fixed by
$\mu^2$ (which is assumed to have the right sign for spontaneous
symmetry breaking) and $\lambda$, $v=[\mu^2/(2\lambda)]^{\frac 1
2}=246$ GeV, and where the covariant derivative is

\be D_\mu\Phi = ( \partial_\mu \Phi + i\frac g 2 {\bar \tau} \cdot
{\bf W}_\mu\Phi- i\frac {g'} 2 B_\mu\Phi\tau_3 ) . \label{Dmu}\ee
(The $\tau_3$ matrix shows explicitly\cite{Willen} in this matrix
representation that the right handed gauge group  is only $U(1)$).
The vacuum value of the $\Phi$ field is

\be < \Phi >  = \frac 1 {\sqrt 2} \left(
\begin{array}{cc}
v & 0  \\
0& \ v \\
        \end{array}\right ) .  \label{VEV}   \ee

 The Lagrangian is, as well known, invariant under the gauge symmetry
 $SU(2)_L$ $\times U(1)_Y$
($\Phi\to L\Phi$ and $\Phi\to \Phi e^{-\frac 1 2 \tau_3\varphi
}$), which is broken spontaneously down to $U(1)_{EM}$ by the $v$
of eq.(\ref{VEV}) . But, in the limit of $g' \to 0$, i.e.
$\theta_W\to 0$, (or if one disregards the gauging) there is, in
fact, also a global $SU(2)_L\times SU(2)_R$ ($\Phi\to L\Phi R$).
Thus in this limit the electro-weak Higgs sector has a global
$SU(2)_L\times SU(2)_R$ symmetry, which is spontaneously broken by
$v$ to $SU(2)_{L+R}=SU(2)_V$ with $W$ and $Z$ degenerate. This
symmetry is usually called the custodial symmetry. In our
discussion we treat this as an exact global symmetry, as the $g'$
term (or the isospin breaking) is not essential here.

This is just like the simplest \lsm\ (without isospin breaking)
for the pion  and $\sigma$. The  analogue of $\Phi$ is

\be \hat\Phi =\frac 1 {\sqrt 2}\left(
\begin{array}{cc}
\sigma+i\pi^0 & i(\pi_1+i\pi_2)  \\
-i(\pi_1-i\pi_2)& \sigma -i\pi^0 \\
        \end{array}\right )=\frac 1 {\sqrt 2} (\sigma \cdot
{\bf 1}+i\bar \pi\cdot \bar \tau ) .  \label{phihat}    \ee

Thus $\phi^0$ corresponds to $(\sigma -i\pi^0)$  and  $\phi^+$ to
$i(\pi_1+i\pi_2)$ = $i\sqrt 2 \pi^+$ in conventional notation.
(The generalization of eq.(\ref{phihat}) to full scalar and
pseudoscalar nonets, $s_k,p_k,\ k=$ 0 to 8,  is $\hat \Phi
=\propto \sum_k (s_k+ip_k)\lambda_k$, where $\lambda_k$ are
Gell-Mann matrices.) The gauged \lsm\ is apart from different
constants identical  to eq.(\ref{higgs})

 \be
 {\cal L}_{L\sigma M} (\hat \Phi ) =  {\rm Tr}
[(D_\mu\hat\Phi )^\dagger D_\mu\hat\Phi] +\hat\mu^2{\rm Tr}
[\hat\Phi^\dagger \hat\Phi] -\hat\lambda {\rm
Tr}[\hat\Phi^\dagger\hat\Phi ]^2, \ee from which $\hat v$ is given
by $\hat v=[\hat \mu^2/(2\hat \lambda)]^{\frac 1 2}=92.6 $ MeV.
(With a $\sigma$ mass of $\approx 600$ MeV  $\hat \lambda
=[m_\sigma/(2\hat v )]^2 \approx 10$, but we actually do not need
that number here).

Now add the two Lagrangians with a small mixing term between the
$\Phi$ and the $\hat \Phi$ proportional to $\epsilon^2$:

\be
 {\cal L} = {\cal L}_{Higgs} ( \Phi
 )  + {\cal L}_{L\sigma M} (\hat \Phi ) +
 \epsilon^2 [{\rm Tr}(\Phi^\dagger\hat \Phi ) +h.c.]/2. \label{Ltot}
\ee This is, in fact,  similar to  the Two-Higgs-Doublet
Model\cite{Hunting}, although the application here is different
and much more down to earth, i.e., not the usual  supersymmetric
or technicolor extensions of the standard model\footnote{A similar
effective Lagrangian as in eq.(\ref{Ltot}) was also suggested
previously\cite{CloseNT}, but with a quite different
phenomenological application in mind, which involved two light
scalar meson nonets and only strong interactions.}.

 The $\epsilon^2$ term breaks a {\it relative} (global)
 $SU(2)_L\times SU(2)_R$ in the sense that
if only one of the two $\Phi$'s is transformed by, say, a left
handed rotation ($L\Phi$ or $L^{'}\hat \Phi$), the relative
symmetry is broken. One could, of course, write down many other
terms\cite{Hunting}, which similarly break another  relative
symmetry\footnote{E.g. a small term $\delta^2Tr (\Phi^\dagger \hat
\Phi \tau_3)$, which breaks  $SU(2)_R$ (c.f. the $g'$ term in
\ref{Dmu})  would be allowed by the overall $SU(2)_L\times
U(1)$.}. But, for our discussion here the simplest possible choice
as in eq.(\ref{Ltot}) is sufficient.

Neglecting for a moment the gauging, and with  our choice of the
$\epsilon$ term the Lagrangian (\ref{Ltot}) has an overall {\it
global} $SU(2)_L\times SU(2)_R$ symmetry (i.e. when $\Phi\to L\Phi
R$ is transformed simultaneously as $\hat\Phi\to L\hat \Phi R$
with the same $L$ and $R$). Also the overall $SU(2)_L\times U(1)$
gauge symmetry is left intact. Let us first discuss how the mixing
term breaks the relative global symmetry, neglecting the gauging.

For $\epsilon=0$ and  $v\ne 0$, $\hat v\ne 0$, there is a triplet
of  Goldstone (or would-be Goldstone) bosons in each sector. With
$\epsilon \ne 0$ one of these triplets, (the pion) gets mass
proportional to $\epsilon$, since the relative symmetry is broken.
The pseudoscalar mass matrix $m^2(0^-)$ gets contributions in two
ways 1) from the fact that the vacuum values are disturbed, and 2)
directly from the mixing term. The corrections to $v$ and $\hat v$
obey the relations:
 \be v^2(\epsilon)=v^2(0)+\epsilon^2\frac {\hat v(\epsilon)}{2 v(\epsilon)\lambda}, \hskip
1cm  \hat v^2(\epsilon)=\hat v^2(0)+\epsilon^2\frac {
v(\epsilon)}{2 \hat v(\epsilon)\hat \lambda}.\label{vshift} \ee
The pseudoscalar squared  mass matrix becomes \be m^2 (0^{-})=
{\left(
\begin{array}{cc}
2\lambda v^2(\epsilon)-\mu^2 & -\epsilon^2  \\
-\epsilon^2& 2\hat \lambda \hat v^2(\epsilon)-\hat\mu^2
 \\
        \end{array}
        \right )
        =+\epsilon^2 \left( \begin{array}{cc}
 \frac {\hat v(\epsilon)}{v(\epsilon)}& -1          \\
-1         & \frac{v(\epsilon)}{ \hat v(\epsilon)} \\
        \end{array}
\right )} ,\ee which is diagonalized by a small rotation
$\theta={\small \frac 1 2} \arctan [2\hat v v/ (v^2+\hat
v^2)]\approx \hat v/v =3.764\cdot 10^{-4}$ (we leave out from now
on the small dependence on $\epsilon$ in $v$ and $\hat v$). Thus
the eigenvalues are 0 and $\epsilon^2(v/\hat v+ \hat v/v)\approx
\epsilon^2 v/\hat v$:
 \be \left(
\begin{array}{cc}
c & s  \\
-s& c \\
        \end{array}\right )m^2 (0^{-})\left( \begin{array}{cc}
c & -s  \\
s& c \\
        \end{array}\right )=
        \epsilon^2\frac{v^2+ \hat v^2}{v \hat v}\left( \begin{array}{cc}
0 & 0  \\
0 & 1 \\
        \end{array}\right )\approx \left( \begin{array}{cc}
0 & 0  \\
0 & \epsilon^2 \frac v{\hat v}  \\
        \end{array}\right ).      \ee
(Here $s=\sin\theta$ and $c=\cos\theta$.) Note that the mass
matrix only depends on $v$,$\hat v$ and $\epsilon$, not on
$\lambda$ nor $\hat \lambda$, which are not well known from
experiment as they are related to the Higgs or $\sigma(600)$
masses. Note also that the pseudoscalar which gets mass is the one
which is related to the pion in the \lsm\ sector, (i.e. it is not
the "would-be Goldstone" of the E-W sector. To get the right pion
mass $\epsilon$ should be about 2.70 MeV.) On the other hand the
Higgs and $\sigma$ bare masses and mixings are only very little
affected since the corrections are proportional to the very small
number $3\epsilon^2$.

As the whole Lagrangian(\ref{Ltot}) still has the full
$SU(2)_L\times U(1) $ gauge  symmetry the remaining Goldstone is
turned, in the usual way, into the longitudinal components of the
$W$ and $Z$ bosons. But, these masses get a small contribution
also from the \lsm\ sector \be M_W^2=g^2(v^2+\hat v^2)/4, \hskip
1cm M^2_Z=(g^2+g'^2)(v^2+\hat v^2)/4 .\ee One can also easily see
that the $\epsilon^2$ term mixes the longitudinal W and the pion
in the expected way by the same angle $\theta$.

Thus the pion gets mass from a small breaking of a relative
symmetry between the E-W and strong interactions, and through a
small mixing with  the longitudinal W, described by $\theta
=3.764\cdot 10^{-4}$. Still, the symmetry remains intact, although
spontaneously broken by the vacuum, for the combined strong plus
electro-weak Higgs sectors.

 Our result is in no way in conflict with the usual reference to
 light quark masses as the source of pion mass. In fact, it is
 natural to assume that the $\epsilon^2$ term arises because of  quantum
 corrections involving virtual quark loops ($\hat \Phi\to q\bar q  \to  \Phi$ or
 even $\hat \Phi\to q\bar q  \to VV\to \Phi$, which would involve  ABJ
 anomaly graphs). In principle, such terms should be calculable
 from QCD+EW gauge theory.

A few concluding  final remarks are in order.  I discussed  a way
of how to look at a certain kind of symmetry breaking, which I
named "relative symmetry breaking".  By mixing two models with the
same symmetry group $SU(2)_L\times SU(2)_R$, one for the scalar
sector of strong interactions, and the other for the Higgs sector
of electro-weak interactions, a relative symmetry is broken giving
in our example rise to a pion  mass, although for the whole model
the chiral symmetry is still exact and unbroken in the Lagrangian,
except for spontaneous symmetry breaking in the vacuum. I believe
that this as a general idea  has not received proper attention in
the literature. It need not be limited to the application
discussed here.

For example, it can be applied to the perhaps simplest spontaneous
symmetry breaking one can think of, - that of O(2) or $U(1)$
symmetry models. One needs two $U(1)$ models, each described by a
"Mexican hat" potential, which are coupled to each other by a
mixing term. In this system of "coupled spontaneous symmetry
breaking" the coupled potentials are "tilted" compared to each
other. Therefore, because of the mixing, one combination of the
two would-be Goldstones gets mass. The whole system would still
have one massless Goldstone, which is a linear combination of the
original massless states. Then, if one gauges the whole system
also this massless scalar is "eaten" by the vector boson, and no
massless states remain.

In this simple $U(1)$ example,  our relative symmetry breaking may
seem trivial, almost self-evident, but its possible generalization
within more complicated structures is often missed in current
literature. Our main conjecture is thus that  a symmetry of the
standard model for strong {\it and} electro-weak interactions
combined can remain exact in the total Lagrangian, although the
symmetry looks "as if it were broken explicitly by one of the
interactions". If this conjecture turns out to be  true it should
have consequences for a better understanding of, say, PCAC, chiral
symmetry and isospin breaking.

\section{Acknowledgements}
 I thank Prof. Shou-hua Zhu for a comment during my recent talk at Peking university that
 this conjecture may lead to calculable radiative corrections to e.g. standard model
 Higgs mass estimates. Support  from EU RTN Contract  CT2002-0311 is gratefully
acknowledged.


\begin{thebibliography}{99}
\bibitem{LSM} J. Schwinger, Ann. Phys. {\bf 2} (1957) 407;
M. Gell–-Mann and M. Levy, Nuovo Cim. XVI
(1960) 705; B. W. Lee, Nucl. Phys. {\bf B9} (1969) 64.
\bibitem{Willen} S. Willenbrock,  "Symmetries of the standard
model", (Lectures given at Theoretical Advance Study Institute in
Elementary Particle Physics (TASI 2004) in  Boulder, Colorado, 6
Jun - 2 Jul 2004), hep-ph/0410370; See also M. Veltman,
Reflections on the Higgs system, (Lectures at the Academic
Training Program at CERN) CERN Yellow report 97-05.
\bibitem{Hunting} See e.g. J. F.  Gunion et al. "The Higgs Hunter's Guide",
(Frontiers in Physics), Addison-Wesley, Reading, MA,  1990,  page
191.
\bibitem{CloseNT} F. E. Close and N. A. T\"ornqvist,
Journal of Physics G (Nucl. Part. Phys.) {\bf 28} (2002)
R249-R267, (hep-ph/0204205); N. A. T\"ornqvist hep-ph/0204215.
(unpublished); D. Black, A. Fariborz, S. Moussa, S. Nasri, J.
Schechter, Phys. Rev. {\bf D64}, (2001) 014031; M. Napsuciale and
S. Rodriguez,  Phys. Rev. {\bf D70} (2004) 094043,
(hep-ph/0407037).
\end{thebibliography}
\end{document}